\documentclass[structabstract]{aa}
\usepackage{natbib}
\usepackage{fixltx2e}
\usepackage[table]{xcolor}
\usepackage{multirow}
\usepackage{graphicx}
\usepackage{txfonts}
\usepackage{tabularx}
\usepackage{longtable}
\usepackage{lscape}
\usepackage{authblk}
\usepackage{url}

\newcommand{\fermi}{${\it Fermi~}$}
\bibpunct{(}{)}{;}{a}{}{,}
\begin{document}
\renewcommand{\UrlBreaks}{\do\/\do\a\do\b\do\c\do\d\do\e\do\f\do\g\do\h\do\i\do\j\do\k\do\l\do\m\do\n\do\o\do\p\do\q\do\r\do\s\do\t\do\u\do\v\do\w\do\x\do\y\do\z\do\A\do\B\do\C\do\D\do\E\do\F\do\G\do\H\do\I\do\J\do\K\do\L\do\M\do\N\do\O\do\P\do\Q\do\R\do\S\do\T\do\U\do\V\do\W\do\X\do\Y\do\Z}

\title{\textit{Fermi}-LAT upper limits on gamma-ray emission from colliding wind binaries}
\subtitle{}
\author[1]{M. ~Werner\thanks{E-mail: Michael.Werner@uibk.ac.at}}
\author[1,2]{O. ~Reimer}
\author[1,2]{A. ~Reimer}
\affil[1]{Institut f\"ur Astro- und Teilchenphysik and Institut f\"ur Theoretische Physik, Leopold-Franzens-Universit\"at Innsbruck, A-6020 Innsbruck, Austria}
\affil[2]{Kavli Institute for Particle Astrophysics and Cosmology, Department of Physics and SLAC National Accelerator Laboratory, Stanford University, Stanford, CA 94305, USA}
\author[1]{K. ~Egberts}
\authorrunning{M. ~Werner et al.}
\institute{}
\date{accepted ... \\ \bf Version: Final Version -- June 4, 2013 }

\abstract
{Colliding wind binaries (CWBs) are thought to give rise to a plethora of physical processes including acceleration and interaction of relativistic particles. Observation of synchrotron radiation in the radio band confirms there is a relativistic electron population in CWBs. Accordingly, CWBs have been suspected sources of high-energy $\gamma$-ray emission since the COS-B era. Theoretical models exist that characterize the underlying physical processes leading to particle acceleration and quantitatively predict the non-thermal energy emission observable at Earth.}
{We strive to find evidence of $\gamma$-ray emission from a sample of seven CWB systems: WR 11, WR 70, WR 125, WR 137, WR 140, WR 146, and WR 147. Theoretical modelling identified these systems as the most favourable candidates for emitting $\gamma$-rays. We make a comparison with existing $\gamma$-ray flux predictions and investigate possible constraints.}
{We used 24 months of data from the Large Area Telescope (LAT) on-board the \textit{Fermi} Gamma Ray Space Telescope to perform a dedicated likelihood analysis of CWBs in the LAT energy range.}
{We find no evidence of $\gamma$-ray emission from any of the studied CWB systems and determine corresponding flux upper limits. For some CWBs the interplay of orbital and stellar parameters renders the \fermi-LAT data not sensitive enough to constrain the parameter space of the emission models. In the cases of WR140 and WR147, the \fermi-LAT upper limits appear to rule out some model predictions entirely and constrain theoretical models over a significant parameter space. A comparison of our findings to the CWB $\eta$ Car is made.}
{}


 \maketitle

\section{Introduction}

Wolf-Rayet stars (WR stars) are hot ($\mathrm{T} >30000 \, \mathrm{K}$), luminous ($\mathrm{L} > 10^{5} \, \mbox{-} \, 10^{6} \, \mathrm{L}_{\odot}$), and massive ($\mathrm{M} >10 \, \mathrm{M}_{\odot}$) evolved stars. They possess strong stellar winds with high mass-loss rates ($\dot \mathrm{M}_{wr} \approx 10^{-5} \, \mbox{-} \, 10^{-4} \, \mathrm{M}_{\odot} \, \mathrm{yr}^{-1}$) and  high terminal velocities ($\mathrm{v}_{\infty} \approx \, 3 \times 10^{2} \, \mbox{-} \, 6 \times 10^{3} \, \mathrm{km} \, \mathrm{s}^{-1}$). The current WR catalogue \citep{vanderhucht,catannex} contains 227 stars, 89 of which are in binary systems. The total Galactic population of WR stars is believed to be in the thousands. For a comprehensive review of WR stars and their properties, see \citet{crowther2007}.\\
A binary system in which the supersonic stellar winds of the WR star and its companion OB-type star collide is called a colliding wind binary (CWB). The collision of the winds leads to the formation of a contact discontinuity at which the ram pressures of the two stellar winds balance out causing two shocks to form. The volume enclosed by these shocks is often referred to as the wind-collision region (WCR). At the shocks particles can be accelerated to relativistic energies, giving rise to a non-thermal particle population. As a consequence non-thermal radiation is emitted from the WCR region \citep{eichler1993, Romero2003, Pittard2006, Reimer2006}.\\
Synchrotron radiation detected from CWBs \citep{Abbott, Chapman} provided the first convincing observational evidence of a relativistic electron population. High-resolution multi-wavelength observations of WR 146 \citep{dougherty1,Connor} and WR 147 \citep{Williams} unequivocally identified the WCR as the origin of the non-thermal radio emission. In the case of WR 140, orbital modulation of the non-thermal synchrotron radiation was detected \citep{eichler1993, white1995,dougherty2}, indicating that the conditions for non-thermal radio emission in the WCR change significantly during the course of an orbit.\\
Theoretical models (e.g. \citealt{eichler1993, Romero2003, dougherty3, Pittard2006, Reimer2006}) that aim to understand the physical processes leading to non-thermal radiation in CWBs also predict the emission of $\gamma$-rays through inverse Compton (IC) scattering, bremsstrahlung, and $\pi^{0}$-decay. Therefore, CWBs may be a yet undetected source class of $\gamma$-ray binaries that do not contain a compact object. Unfortunately, most values of the relevant input parameters for modelling (e.g. orbital parameters, acceleration mechanism, acceleration efficiency, stellar surface magnetic field strength, and the radiation fields) are characterized by large observational uncertainties or, even worse, are presently unknown. Consequently, the magnitude of the predicted $\gamma$-ray flux, the spectral shape, as well as the contributions of the principally responsible physical processes, can vary considerably. By studying population aspects \citet{Reimer2007,ReimerAIP2009} show that for a maximum acceleration efficiency the binary separation, the distance to Earth, and the kinetic power of the stellar winds have the strongest influence on $\gamma$-ray flux. For large binary separations the dominant $\gamma$-ray emission process was found to be IC-scattering \citep{Reimer2006}.\\
Convincing observational evidence of $\gamma$-ray emission from CWBs does not exist yet (with the notable exception of $\eta$ Car). The COS-B mission detected four $\gamma$-ray sources that are spatially consistent with the location of WR 140, WR 125, WR 98, WR 105 considering the instrument's source localization capability \citep{pollock1987}. Observations by the EGRET instrument in the 100 MeV-10 GeV energy range revealed an unidentified point-like $\gamma$-ray source, 3EG~J2022+4317 in positional coincidence with WR 140 \citep{Romero1999}. Very high-energy $\gamma$-ray observations (80 GeV - 10 TeV) of WR 146 and WR 147 by the MAGIC telescope yielded upper limits only \citep{Alui}. Observations by AGILE \citep{agile} and \fermi-LAT \citep{Abdo2010} detected a $\gamma$-ray source compatible with the spatial location of $\eta$ Car. Follow-up studies by \citet{klaus} revealed variability that is correlated with the orbital motion of $\eta$ Car, providing strong evidence that the $\gamma$-ray source is indeed associated with the $\eta$ Car system. However, an apparently unique system like $\eta$ Car cannot be considered a typical representative for the CWB population \citep{parkin}.\\
Launched on 11 June 2008, the Large Area Telescope (LAT) on-board the \textit{Fermi} Gamma-ray Space Telescope is the most sensitive $\gamma$-ray telescope to date, covering an energy range from 20 MeV to 300 GeV. A population study of known CWBs \citep{Reimer2007,ReimerAIP2009} identified systems in which the predicted $\gamma$-ray flux is above the pre-launch predicted five-year sensitivity ($2 \times 10^{-8} \, \mathrm{ph} \, \mathrm{cm}^{-2} \, \mathrm{s}^{-1}$) of the LAT for sources located in the Galactic plane. Seven CWBs were considered as being favourable to detection by the LAT. With the exception of WR 11 (which is the closest known CWB), these CWBs have large binary separations and therefore long ($>$2000 days) orbital periods, a configuration that is believed to yield favourable conditions for detection. A detailed study of WR 140, WR 146, and WR 147 has been carried out using known orbital and stellar parameters \citep{Reimer2006, Reimer2009}. An overview of the stellar and physical characteristics of the analysed CWB systems is given in Table \ref{wrparam}. We studied this sample using 24 months of \fermi-LAT data and present the results here.

\section{Observation, data analysis, and results}
\label{data}

\subsection{Observation}
\label{observation}
On 4 August 2008, the \fermi-LAT began regular science observations that allow it to cover the entire sky roughly every three hours. A review of the \fermi-LAT instrument and performance is given in \citet{lat}. The analysed dataset covers 24 months, starting on 4 August 2008 and ending 1 August 2010, which is the same time period as covered by the \textit{Fermi} Large Area Telescope Second Source (2FGL) Catalog \citep{2fgl}. Data reduction and analysis was performed using the \textsc{Fermi Science Tools v9r23} software package\footnote{For details visit the Fermi Science Support Center (FSSC): \texttt{http://fermi.gsfc.nasa.gov/ssc/}}.\\
Only events from the Pass 7 ``Source'' event class have been selected to obtain high-quality photon data. Photons emanating from the Earth's limb were rejected by excluding time intervals in which any part of the analysed region was observed at zenith angles greater than $100^{\circ}$ and the observatory's rocking angle was greater than $52^{\circ}$. For each analysed CWB, the remaining photons with an energy E $>$ 95 MeV and originating within a square of $21.2^{\circ}$ edge length (called the region of interest, ROI) centred on the nominal position of the binary system were used for analysis.

\subsection{Maximum likelihood analysis}
\label{likelihood}
All analysed CWBs are located close to the prominent Galactic plane, where diffuse emission is the dominating source of $\gamma$-rays. Diffuse emission arises from interactions of cosmic rays with the interstellar medium and radiation fields as they propagate in the Milky Way. Additionally, an isotropic $\gamma$-ray background is present that is due to extragalactic diffuse $\gamma$-rays and the remaining residual (misclassified) cosmic rays. Therefore, detecting faint point sources at the detection threshold on top of a bright diffuse emission signal is one of the main challenges of the presented analysis, so meticulous modelling of the diffuse $\gamma$-ray emission is necessary. We used the \emph{gal$\_$2yearp7v6$\_$v0.fits} model for the Galactic diffuse emission in combination with the \emph{iso$\_$p7v6source.txt} model for the isotropic background\footnote{More details about the background  models, their use, and the corresponding files are available from\\ \url{http://fermi.gsfc.nasa.gov/ssc/data/access/lat/BackgroundModels.html}}. This is consistent with the 2FGL catalogue. 
Using the tool \texttt{gtlike}, binned likelihood analysis is applied to the selected data using the ``P7SOURCE\_V6'' instrument response functions \citep{pass7}. The covered energy range 95.6 MeV $<$ E $<$ 44.9 GeV is divided into 19 logarithmically spaced energy bins. This results in energy bins that coincide with the energy planes of the used model for the Galactic diffuse emission. For each analysed CWB the source model (the null hypothesis) used in the likelihood analysis includes all sources listed in the 2FGL catalogue contained within a circle of $15^{\circ}$ radius centred on the CWBs nominal position.
The majority of sources are modelled using a power law
\begin{equation} \frac{\mathrm{d}N}{\mathrm{d}E}=A\:E^{-\Gamma}\end{equation}
where $\frac{\mathrm{d}N}{\mathrm{d}E}$ is the differential flux (given in units of $\mathrm{cm}^{-2} \, \mathrm{s}^{-1} \, \mathrm{MeV}^{-1}$), $A$ a normalization factor, and $\Gamma$ the spectral index. In case a log-parabolic model function provides a better fit,
\begin{equation} \frac{\mathrm{d}N}{\mathrm{d}E}=A \left( \frac{E}{E_b} \right)  ^{-(\alpha+\beta log(\frac{E}{E_b}))}\end{equation} is used, which introduces a break energy $E_{b}$, spectral index $\alpha$, and curvature index $\beta$ as additional parameters. Known $\gamma$-ray pulsars were modelled by an exponentially cut-off power law
\begin{equation} \frac{\mathrm{d}N}{\mathrm{d}E}=A\: E^{-\Gamma}\mathrm{exp} \left[ -\frac{E}{E_{\mathrm{cutoff}}} \right] \end{equation}
where $E_{\mathrm{cutoff}}$ is the cut-off energy, $A$ a normalization factor, and $\Gamma$ the spectral index. Extended sources (e.g. Vela X, MSH 15-52, W51C, and the Cygnus Loop) were modelled by the same spectral and spatial models used in the 2FGL catalogue. Finally, the considered CWBs were modelled using a power law, and their positions were fixed at the coordinates listed in Table \ref{wrparam}. The spectral parameters of all sources that have an angular separation of less than or equal to $7^{\circ}$ from the CWB are left free to be fitted by the likelihood analysis, while those with an angular separation greater than $7^{\circ}$ are fixed to the values listed in the 2FGL catalogue. In the likelihood analysis the ``Prefactor'' of the Galactic diffuse model and the ``Normalization'' of the isotropic model are free parameters. The total number of free parameters in each source model used in the analysis is given in Table \ref{analysis}.\\
Source detection significance at a specified position can be statistically expressed using the test statistic (TS) value
\begin{equation}
TS=-2\ln(L_{max,0}/L_{max,1})
\end{equation}
where $L_{max,0}$ is the maximum likelihood value for a model without an additional source (the null hypothesis), and $L_{max,1}$ is the maximum likelihood value for a model that includes an additional source at a specified location. Both $L_{max,0}$ and $L_{max,1}$ are determined using the maximum likelihood method. The influences of other sources and of the background are thereby taken into account. In the limit of large counts the TS value follows a $\chi^2$-distribution \citep{mattox} and is equal to the square of the detection significance $\sigma$ for a point source.
The criterion for considering sources for inclusion in the 2FGL was chosen to be $TS > 25$ \citep{2fgl}. However, for sources in the Galactic plane (e.g. magnetars \citealt{magnetars}) this was found not to be sufficiently conservative, mainly due to the large uncertainties in the Galactic diffuse background model \citep{strong_diffuse2}. Systematic uncertainties resulting from source confusion caused by the high number density of point-like and extended sources in the Galactic plane, as well as additional uncertainties introduced by the existence of sub-threshold sources (which we estimate to be as high as 10\%), complicate source detection further. Identification of a source with a specific astrophysical object not only requires spatial coincidence but also observation of a characteristic quantity (such as orbital modulation, periodicity, and correlated variability) that is intrinsically linked to the object under consideration.
A summary of the results obtained using likelihood analysis of the CWBs is given in Table \ref{analysis}.
\begin{table*}
\caption{Likelihood analysis results from the sample of studied CWBs}
\label{analysis}
\centering
\renewcommand{\arraystretch}{1.5}
\begin{tabular}{p{3cm} c c c c c c c c}
\hline\hline
Parameter									&Unit			&WR 11					&WR 70			&WR 125					&WR 137					&WR 140					&WR 146		&WR 147\\
\hline
Free parameters \newline in source model	&			&25  					&45				&32						&71						&71						&77			&69\\
TS					&		&0.0 &0.0 &0.0 &5.9 &1.8 &5.3 &4.6\\
\end{tabular}

\end{table*} The likelihood analysis did not reveal any excess above detection threshold for any of the analysed CWBs. No point sources were detected that are compatible with the nominal positions of the CWB sample. The CWBs located in the Cygnus region have the highest TS values, because the Cygnus region is one of the most complex regions in the $\gamma$-ray sky, displaying an excess of $\gamma$-rays on both large and small scales that is not adequately modelled in the standard Galactic diffuse background model chosen in this analysis \citep{luigi_cygnus}.

\subsection{Upper limit determination}
\label{upperlimit}
Upper limits presented in this paper were calculated using the method described by \citet{fc}. Assuming Poisson statistics, this method uses the number of mean signal counts and mean background counts to construct confidence intervals. The observed signal counts are extracted using the likelihood analysis software. Background count extraction, however, warrants a closer look and is described in the following.
The number of mean background counts $C^{bg}$ consists of two components: observed counts attributed to the isotropic $\gamma$-ray emission and those attributed to the Galactic diffuse $\gamma$-ray emission. In the applied binned likelihood analysis technique, the number of mean background counts is given by \begin{equation} C^{bg} = \sum_{i} C_{i}^{bg} = \sum_{i} C_{i}^{gal} + \sum_{i} C_{i}^{iso}\end{equation}
with $C_{i}^{bg}$ the number of mean background counts, $C_{i}^{gal}$ the number of Galactic diffuse emission counts, and $C_{i}^{iso}$ the number of counts attributed to isotropic emission, per energy bin $i$. For each energy bin $i$, the average differential intensity $I_{i}$ of the background model \emph{gal$\_$2yearp7v6$\_$v0.fits} covered by the 95\% containment angle of the point spread function (PSF) is determined. The resulting average differential intensity $I_{i}$ is converted to an average flux $F_{i}^{bg}$ of the Galactic diffuse emission using
\begin{equation}F_{i}^{bg} = N_{gal} \: I_{i} \: \left[ E_{i}^{u} - E_{i}^{l} \right] \: A^{psf}_{i} \end{equation} 
wherein $N_{gal}$ is the normalization of the \emph{gal$\_$2yearp7v6$\_$v0.fits} (obtained using likelihood analysis), $A^{psf}_{i}$ is the area covered by the 95\% containment angle of the PSF in energy bin $i$, $E_{i}^{u}$ the upper and  $E_{i}^{l}$ the lower bound of energy bin $i$. The number of Galactic diffuse background counts $C^{gal}_{i}$, contained within the 95\% containment angle of the PSF, is then computed using the exposure $V_{i}^{exp}$ of each energy bin $i$
\begin{equation}C^{gal}_{i} = F_{i}^{bg} \: V_{i}^{exp} \: \mbox{.}\end{equation}
The number of counts from the isotropic emission component $C_{i}^{iso}$ per energy bin $i$ is determined using 
\begin{equation}C_{i}^{iso} = \frac{C_{tot}^{iso}}{(\mbox{ROI})^{2}} \: A^{psf}_{i}\end{equation}
wherein $C_{tot}^{iso}$ is the total number of isotropic background counts in the energy bin $i$ (given by the likelihood fit of the \emph{iso$\_$p7v6source.txt} model), $\mbox{ROI}$ is the size of the region of interest (see \ref{observation}) used in the likelihood analysis, and $C^{gal}_{i}$ and $C_{i}^{iso}$ are the quantities required to determine the number of mean background counts $C_{i}^{bg}$. Using the Feldman and Cousins methodology 95\%-confidence upper limits $C_{i}^{ul}$ were calculated and subsequently converted into 95\%-confidence upper limits on the flux $F_{i}^{ul}$
\begin{equation}F_{i}^{ul} = \frac{C_{i}^{ul}}{ V_{i}^{exp}} \: \mbox{.}\end{equation}
For each CWB the flux upper limits and their corresponding energy bins are listed in Table \ref{ul}. Our upper limit calculation only includes statistical uncertainties.
\begin{table*}
\caption{2-$\sigma$ upper limits on $\gamma$-ray flux from the analysed CWB sample.}
\label{ul}
\centering
\renewcommand{\arraystretch}{1.5}
\begin{tabular}{l| c c c c c c c c}
\hline\hline
Energy bin		 	&Unit						&WR 11					&WR 70			&WR 125					&WR 137					&WR 140					&WR 146		&WR 147\\
MeV					& $\mathrm{ph} \, \mathrm{cm}^{-2} \, \mathrm{s}^{-1}$			 			& 						& 				&	 					& 						& 						& 			&\\
\hline
95.6 - 132.1			&			$10^{-8}$		&2.7 &4.6 &3.6 &4.2 &3.7 &4.3 &4.3\\
132.1 - 182.6		&							&1.5 &2.7 &2.2 &2.6 &2.3 &2.8 &2.8\\
182.6 - 252.5		&							&0.9 &1.7 &1.5 &1.7 &1.5 &1.9 &1.8\\
252.5 - 349.1		&							&0.5 &1.2 &1.0 &1.1 &1.0 &1.3 &1.3\\
349.1 - 482.6		&							&0.3 &0.8 &0.7 &0.7 &0.6 &0.9 & 0.9\\
482.6 - 667.2		&			$10^{-9}$		&2.1 &5.1 &4.4 &4.7 &4.2 &5.8 &5.8\\
667.2 - 922.4		&							&1.3 &3.1 &2.8 &3.1 &2.7 &3.9 &3.9\\
922.4 - 1275.2		&							&0.9 &1.9 &1.8 &2.0 &1.7 &2.5 &2.5\\
1275.2 - 1763.0		&							&0.5 &1.2 &1.1 &1.2 &1.1 &1.6 &1.6\\
1763.0 - 2437.3		&			$10^{-10}$		&3.4 &7.3 &6.8 &7.6 &6.5 &10.0 &9.9\\
2437.3 - 3369.5		&							&2.2 &4.7 &4.2 &4.9 &4.1 &6.3 &6.2\\
3369.5 - 4658.3		&							&1.5 &3.1 &2.7 &3.1 &2.7 &4.1 &4.0\\
4658.3 - 6440.1		&							&1.1 &2.1 &1.9 &2.2 &1.8 &2.7 &2.7\\
6440.1 - 8903.3		&							&0.7 &1.5 &1.3 &1.5 &1.3 &2.0 &2.0\\
8903.3 - 12308.8		&							&0.6 &1.3 &1.0 &1.7 &1.0 &1.5 &1.5\\
12308.8 - 17016.7	&			$10^{-11}$		&4.1 &9.3 &7.7 &8.5 &7.6 &11.1&11.1\\
17016.7 - 23525.4	&							&4.4 &6.4 &6.3 &6.9 &5.3 &8.0 &8.0\\
23525.4 - 32523.6	&							&4.7 &7.4 &5.6 &5.8 &6.1 &6.1 &6.1\\
32523.6 - 44963.5	&							&4.8 &4.6 &3.5 &3.7 &3.8 &5.2 &5.2\\
\hline
95.6	 - 44963.5		&			$10^{-9}$		&3.0	 &5.5 &3.5 &13.7&9.6 &16.2&17.9\\
&	& & & & & & &\\
\end{tabular}
\end{table*}

\section{Comparison with theoretical modelling}
\label{theo}
We compare the upper limits obtained in Section \ref{upperlimit} to theoretical models of CWB systems that provide $\gamma$-ray predictions for the CWB system studied. We focus on the following three models. The model published in \citet{Romero2003} calculates $\gamma$-ray flux due to IC scattering, pion decay, and relativistic bremsstrahlung. Underlying relativistic particle distribution are calculated using analytical approximations for the involved acceleration effects and energy loss mechanisms. Orbital dynamics, propagation effects, IC scattering in the Klein-Nishina regime, and anisotropy of the scattering process, as well as photon-photon pair production, are not taken into account. Predictions for the $\gamma$-ray emission are given for WR 140, WR 146 and WR 147.\\
The emission model for WR 140 by \citet{Pittard2006} is based on 2D hydrodynamical simulations of the stellar wind dynamics. These simulations yield temperature and density distributions from which the broadband thermal radiation is derived. Non-thermal electron and ion distributions are generated through diffusive shock acceleration (DSA) in the volume enclosed by the WCR. Propagation effects, orbital dynamics, radiative, and adiabatic energy losses are included. Anisotropy effects of the IC scattering, the Klein-Nishina regime, and photon-photon absorption are not taken into account. Gamma-ray flux predictions are given for a variety of scenarios.\\
The semi-analytical \citet{Reimer2006} emission model for CWBs calculates the non-thermal broadband radiation from the steady-state proton and electron distributions. The diffusion-loss equation considers all relevant radiative losses, adiabatic cooling, and diffusive particle acceleration at the shock out of a pool of thermal wind particles. Spatial diffusion/convection are taken into account using suitable approximations. The IC emissivity and energy losses extending into the Klein-Nishina regime are calculated, and take the anisotropic nature of the scattering process into account in this environment. Pair production due to photon-photon absorption is also evaluated. \citet{Reimer2006} and \citet{Reimer2007,ReimerAIP2009,Reimer2009} predict the gamma-ray emission of selected WR binary systems.
All three models assume single-photon IC scatterings.
\subsection{WR 140}
\begin{figure}[t]
\centering
\includegraphics[width=12cm]{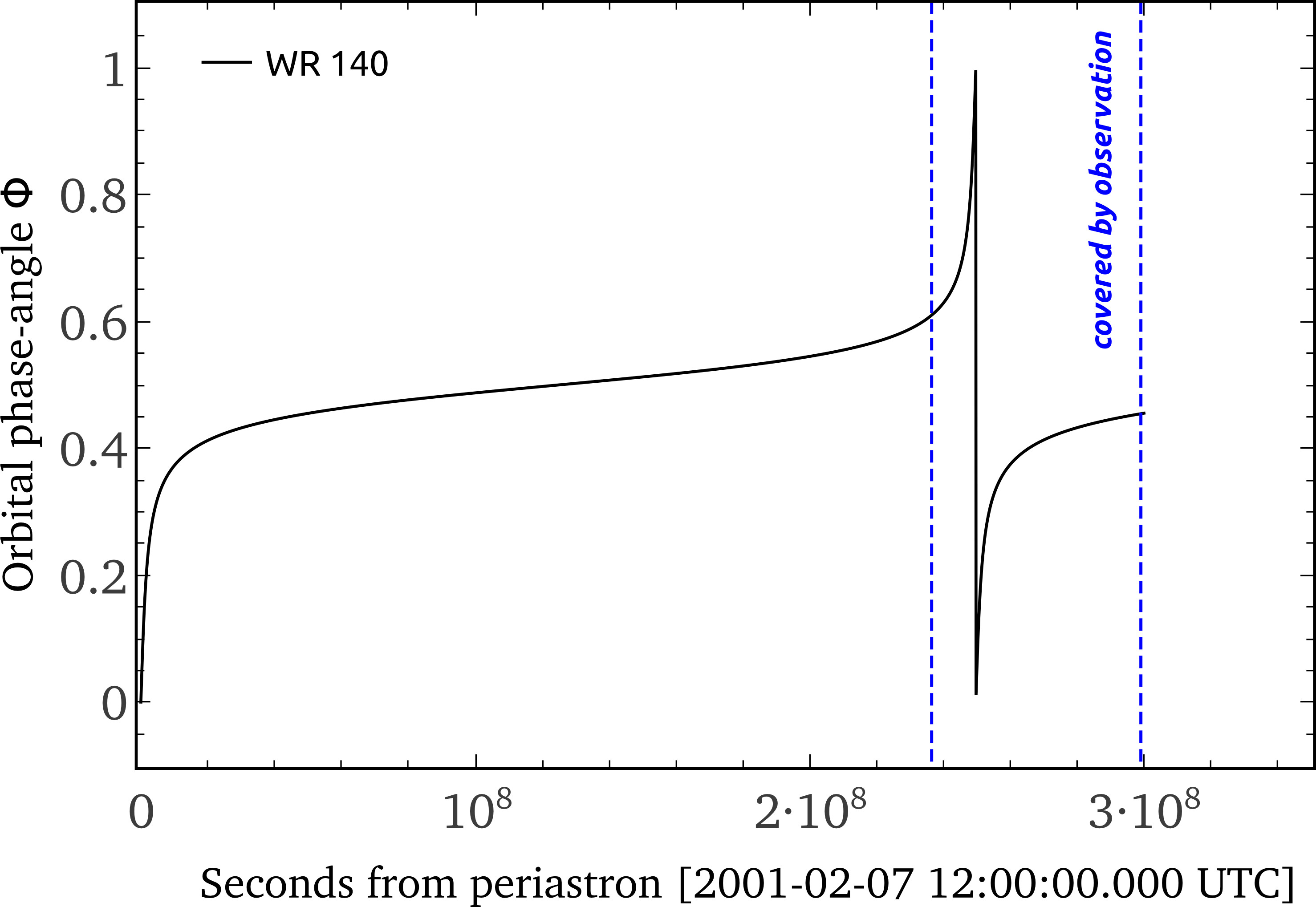}
\caption{Orbital phase angles of WR 140, derived using parameters given in \citet{monnier}. Phase angles covered by the data lie within the two dashed vertical lines. Reference point is the periastron passage on 2001-02-07 12:00:00.000 UTC. \label{fig:phase}}

\end{figure}

\begin{figure}[t]
\centering
\includegraphics[width=12cm]{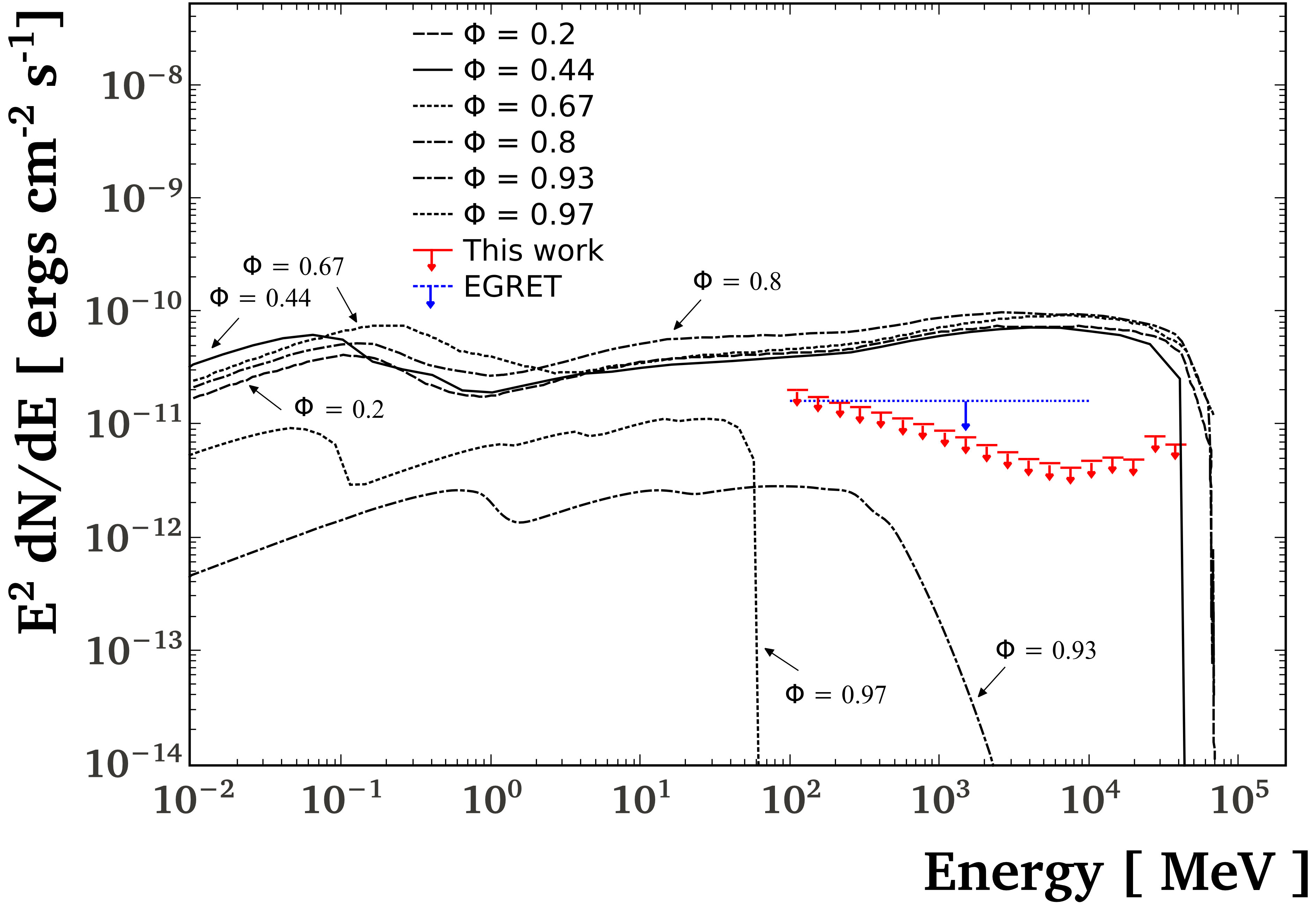}
\caption{2-$\sigma$ differential flux upper limits (\textit{red lines}) of WR 140 compared to theoretical modelling explained in \citet{Reimer2006}. Black curves show IC spectra of WR 140 orbital phase angles $\Phi$ = 0.97 (\textit{bottom dash-dotted line}), $\Phi$ = 0.93 (\textit{bottom dotted line}), $\Phi$ = 0.2 (\textit{dashed line}), $\Phi$ = 0.44 (\textit{solid line}), $\Phi$ = 0.67 (\textit{top dotted line}), and $\Phi$ = 0.8 (\textit{top dash-dotted line}). The EGRET upper limit (\textit{blue dotted line}) is taken from \citet{Reimer2006}. \label{fig:wr140}}

\end{figure}

WR 140 is often considered the archetypal CWB. Unlike most of the CWB systems analysed in this work, the orbital and stellar parameters are known accurately (see Table \ref{wrparam}). This reduces the influence of the uncertainties in the input parameter space of theoretical models significantly, when interpreting the modelling results. We use the $\gamma$-ray flux predictions of \citet{Reimer2006} where a surface magnetic field strength of $100$ G was used. This value leads to equipartition field values at the shock location (see e.g., \citealt{Romero2003}). Because WR 140 is a long-period binary, we assume IC-scattering of the electrons in the radiation field of the OB-companion to be the dominant $\gamma$-ray emission process. 
We note that a realistic evaluation of the $\gamma$-ray emission from IC-scattering in the context of colliding wind binaries must be based on the use of the full Klein-Nishina cross section and take the anisotropic nature of the scattering process into account, in addition to accounting for $\gamma$-ray absorption by pair production \citep{Reimer2006}. This leads to orbital flux variations whose magnitude can be influenced by the geometry of the system. Klein-Nishina and anisotropy effects yield spectral and variability signatures even in systems with circular orbits \citep{cerutti}.
We use the true anomaly, henceforth referred to as the phase angle $\Phi$, as a measure of the orbital configuration. The phase angle at a given time T is defined as the angle between the lines connecting the stars at periastron and at time T. The system passed from orbital phase angle $\Phi \sim 0.61$ through periastron to $\Phi \sim 0.45$ in the time period covered by the data (derived using parameters from \citet{monnier}); see Figure \ref{fig:phase}. Periastron took place on 12 January 2009. The flux upper limits listed in Table \ref{ul} have been converted into $E^{2} \frac{\mathrm{d}N}{\mathrm{d}E}$ flux upper limits, i.e. differential flux scaled by the energy squared, using a power law with spectral index $\Gamma = 2$. A comparison of the differential flux upper limits with the modelling results for different orbital phase angles $\Phi$ is shown in Figure \ref{fig:wr140} along with the EGRET upper limit \citep{Reimer2006}. For orbital phase angles $0.2 < \Phi < 0.8$ our differential flux upper limits are about an order of magnitude below model predictions. For orbital phase angles close to the periastron ($0.2 > \Phi > 0.8$) we cannot place any constraints on predictions as of \citet{Reimer2006}. Because of the high eccentricity of the system the period in which WR 140 passes through these phase angles is very short, corresponding to just 5\% of the observation time of this analysis. In the case of WR 140, the model of \citet{Reimer2006} is disfavoured in the parameter space presented there.\\
Additionally, we adapted (assuming a canonical spectral index $\Gamma = 2$) the $\gamma$-ray flux prediction from \citet{Romero2003} to the energy range covered by our analysis. This gives a value of $F \approx 2.3 \times  10^{-7} \, \mathrm{ph} \, \mathrm{cm}^{-2} \, \mathrm{s}^{-1}$, which is $\sim 24$ larger than the upper limit $F_{ul} \approx 9.6 \times  10^{-9} \, \mathrm{ph} \, \mathrm{cm}^{-2} \, \mathrm{s}^{-1}$ (following the procedure detailed in Section \ref{upperlimit} but with one bin covering the entire energy interval i. e. 95.6 MeV $<$ E $<$ 44.9 GeV) that we obtain for the same energy range. We can exclude the prediction from the model of \citet{Romero2003} for the $\gamma$-ray flux of WR 140 with a statistical confidence of at least 95\%.\\
A comparison with the ten model predictions by \citet{Pittard2006} splits into two categories. Whereas models A-D and F (100 MeV – to 10 GeV) are not constrained by the Fermi upper limit on WR 140, models E and G-J appear to be ruled out. This dichotomy arises due to model parameters common to either one of the two categories. Models that are in conflict with the Fermi data feature high normalization factors of the non-thermal electron density relative to the thermal particle internal density, and low normalization factors of the magnetic field density relative to the thermal particle energy density. Model E, which would mainly range among the group A-F, seems to be excluded by reason of a particularly low value of p, the model's normalization to the radio emission.

\subsection{WR 147}
\begin{figure}[t]
\centering
\includegraphics[width=12cm]{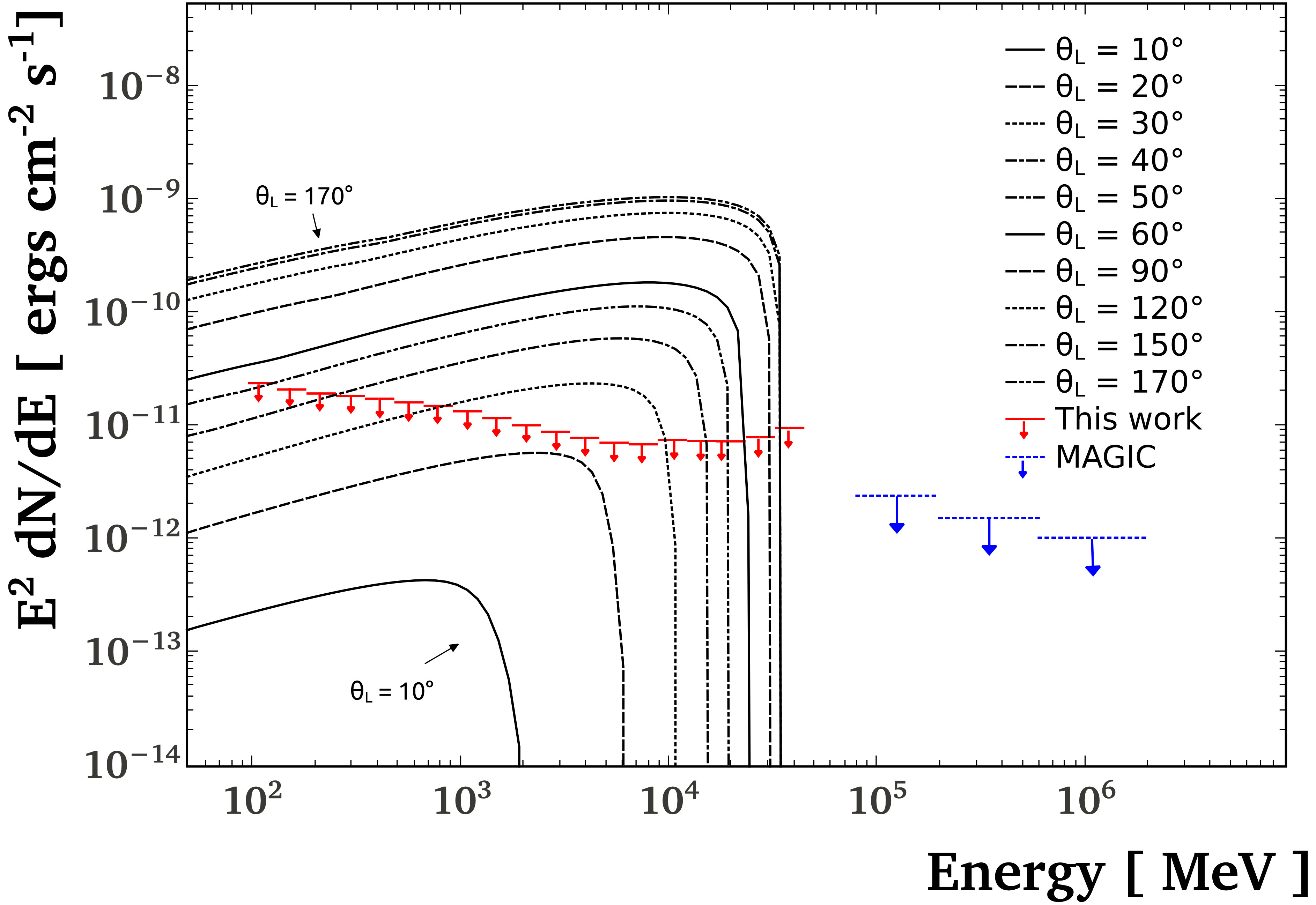}
\caption{2-$\sigma$ differential flux upper limits (\textit{red lines}) of WR 147 compared to modelling explained in \citet{Reimer2009}. The model curves shown here correspond to a parameter set where a 30 G surface magnetic field for the B-star has been used. Black curves show IC spectra of WR 147 for possible geometric viewing angles $\Theta_{L} = 10^{\circ}$, $20^{\circ}$, $30^{\circ}$, $40^{\circ}$, $50^{\circ}$, $60^{\circ}$, $90^{\circ}$, $120^{\circ}$, $150^{\circ}$, and $170^{\circ}$ (\textit{lower to upper curve}). We assume an inclination of $i=85^{\circ}$ and a binary separation of $d \approx 4800$ AU. The MAGIC upper limits (\textit{blue dotted lines}) are taken from \citet{Alui}.\label{fig:wr147}}
\end{figure}

The orbital parameters of the extremely long-period binary WR 147 are unknown (see Table \ref{wrparam}). Model predictions of the expected $\gamma$-ray flux are taken from a dedicated parameter study wherein IC-scattering of the electrons in the radiation field of the OB-companion is identified as the dominant mechanism of $\gamma$-ray emission \citep{Reimer2009}. Here, we assume an inclination angle of $i=85^{\circ}$ that corresponds to a binary separation $d \approx 4800$ AU. X-ray observations around the time of the \fermi-LAT observations indicate that the B-star was likely very close to being in front of the WR star along the sight line \citep{Zhekov2}. This corresponds to the situation adopted in \citet{Reimer2009}. The allowed range of the inclination angle in the case of maximum possible acceleration efficiency, however, differs from the limits given by \citet{Zhekov2}. Using $i=85^{\circ}$ and a surface magnetic field strength of 30 G, 11\% of the B-star's wind kinetic power must be used for electron acceleration at maximum efficiency to account for the observed synchrotron flux \citep{Reimer2009}.\\
A comparison of the differential flux upper limits with model results for different geometrical viewing angles is shown in Figure \ref{fig:wr147}. The geometrical viewing angle is defined as $\cos \Theta_{L} = \cos \psi \sin i$, wherein $i$ is the inclination and $\psi$ is the angle between the projected line of sight and the line connecting the binary component stars. For more details, as well as an illustration of the orbital configuration see \citealt{Reimer2009}. For the assumed model parameters we can exclude all geometrical viewing angles $\Theta_{L} >  20^{\circ}$. Upper limits reported by the MAGIC Collaboration \citep{Alui} have also been included in Figure \ref{fig:wr147}, although they do not any pose additional constraints on the geometric viewing angle for $i=85^{\circ}$. Model predictions for inclination angles of $i < 85^{\circ}$ are solely constrained by the MAGIC and INTEGRAL observations \citep{Reimer2009}. In cases where low inclination angles are favoured (e.g., \citealt{Contreras,dougherty3,Zhekov2}), our derived upper limits may indicate that electron acceleration cannot be as efficient as considered in \citet{Reimer2009}.\\
As in the case of WR 140, we adapted the $\gamma$-ray flux prediction from \citet{Romero2003} to the energy range covered by our analysis by assuming a canonical spectral index $\Gamma = 2$. This gives a value of $F \approx 2.0 \times  10^{-7} \, \mathrm{ph} \, \mathrm{cm}^{-2} \, \mathrm{s}^{-1}$, which is above, by an order of magnitude, the flux upper limit $F_{ul} \approx 1.8 \times  10^{-8} \, \mathrm{ph} \, \mathrm{cm}^{-2} \, \mathrm{s}^{-1}$ from our analysis in the same energy range. We can rule out the $\gamma$-ray flux prediction for WR 147 from the model of \citet{Romero2003} with a statistical confidence of at least 95\%.

\subsection{WR 11, WR 70, WR 125, WR 137, and WR 146 }
For WR 11, WR 70, WR 125, WR 137, and WR 146, the expected $\gamma$-ray flux at Earth has been calculated in the context of a statistical study \citep{Reimer2007,ReimerAIP2009}. Except for WR 11, the orbits of these systems are not known. A circular orbit and an inclination of $i=90^{\circ}$ are used for all systems to avoid systematic effects unknown orbital parameters and to facilitate a better comparison of these systems. The orbital phase angle $\Phi = 0$ then corresponds to the WR star being in front of the OB-star along the line of sight. A stellar surface magnetic field of 100 G and the highest, physically plausible  non-thermal injection power\footnote{The highest possible non-thermal injection power is limited by the wind's particle flux entering the shock, the available wind kinetic power, and the shock acceleration mechanism approaching the Bohm-limit (see e.g., \citealt{Reimer2006}).} is assumed for all CWBs in order to emphasize the influence of the stellar wind parameters on the $\gamma$-ray flux. An $E^{-2}$ power-law electron injection spectrum is used and only contributions to $\gamma$-ray emission from the IC-scattering of electrons in the radiation field of the OB-companion star are taken into account \citep{Reimer2007,ReimerAIP2009}. Within reasonable bounds for the flux normalization set by the total kinetic power of the wind and conservation of particle numbers these assumptions result in the maximum possible $\gamma$-ray flux that can be expected to be observable at Earth. The WR binaries WR70, WR125, WR137, and WR137 have not been detected in the analysed data set, and the determined upper limits (see Table \ref{ul}) do not impose constraints on the emission model that was used for predictions regarding the population of WR binaries by \citet{Reimer2007,ReimerAIP2009}. The $\gamma$-ray flux upper limits of WR 146 obtained from MAGIC observations \citep{Alui} do not impose additional constraints on the emission model of \citet{Reimer2006}.\\ 
The model prediction for the $\gamma$-ray flux of WR 146 by \citet{Romero2003} is adapted to the energy range covered by our analysis by assuming a canonical spectral index $\Gamma = 2$. The thus obtained upper limit ($1.86 \times  10^{-8} \, \mathrm{ph} \, \mathrm{cm}^{-2} \, \mathrm{s}^{-1}$) differs only by 10\% when compared to the flux upper limits presented here ($1.62 \times  10^{-8} \, \mathrm{ph} \, \mathrm{cm}^{-2} \, \mathrm{s}^{-1}$). Because this deviates only mildly from the flux upper limits presented here we conclude that we cannot rule out the model of \citet{Romero2003} in the case of WR 146.

\section{Summary and discussion}
\label{Disc}
Using 24 months of \fermi-LAT data, we find no evidence of $\gamma$-ray emission from any of the analysed CWBs. Since all analysed CWBs are located in the Galactic plane, the analysis is complicated by the intense diffuse emission and uncertainties in the Galactic diffuse emission background model used in the likelihood analysis. At least for energies below 1 GeV, we are clearly in a regime that is background-limited. Upper limits on the observable $\gamma$-ray flux, which are independent of spectral model assumptions, were calculated using the methodology as outlined in Section \ref{ul}. Upper limits are compared to the CWB $\gamma$-ray emission model developed by \citet{Reimer2006}. Accordingly, we consider $\gamma$-ray emission due to IC-scattering of relativistic electrons on the radiation field of the OB-companion star to be the dominating $\gamma$-ray production mechanism.\\
Comparing the results from our analysis with the model predictions we find two categories. The first concerns WR 140 and WR 147. In the case of WR 140, orbital parameters are sufficiently well known. We show that the data is sensitive enough to constrain parameters of the adopted $\gamma$-ray emission model. Our upper limits on the $\gamma$-ray flux are up to an order of magnitude below the model predictions, disfavouring the model of \citet{Reimer2006} in the presented parameter space. In the case of WR 147, the orbital parameters are unknown. However, for the assumed parameter space of the emission model published in \citet{Reimer2009}, the upper limits are sensitive enough to constrain the geometric viewing angle, $\Theta_{L} <  20^{\circ}$. Additionally, we also made a comparison with the $\gamma$-ray flux predictions by \citet{Romero2003} and can exclude them for both WR 140 and WR 147. Similarly, the flux predictions given by \citet{Pittard2006} for models E and G-J are not compatible with the data. The non-detection of WR 11, WR 70, WR 125, WR 137, and WR 146 constitute the second category for which we cannot constrain the theoretical models considered herein.\\
Within reasonable bounds set by the total kinetic power of the stellar wind, conservation of particle numbers and, if known, the orbital as well as stellar parameters of the CWB in question, the model of \citet{Reimer2006} uses parameters that give the maximum $\gamma$-ray flux that can be expected to be observable at Earth. Consequently, the constraints we imposed on the $\gamma$-ray emission model in the case of WR 140 need to be investigated further to determine which subset of the model parameter space is still compatible with the data. Furthermore, the model of \citet{Reimer2006} contains a number of simplifying assumptions that have been made in order to allow for an analytical treatment of particle acceleration and subsequent $\gamma$-ray emission processes occurring in CWB systems. Propagation effects inside the WCR (e.g. convection, diffusion) were taken into account \citep{Reimer2006} and the radiation field of the WR star was neglected owing to the dominance of the OB-star's radiation field density at the location of the wind collision region. Additionally, for many of the studied systems, stellar and orbital parameters have not been determined or have large uncertainties.\\
The values given in Table \ref{wrparam} for mass-loss rate, surface temperature, and the terminal velocity of the stellar wind are determined from observational data that have to be evaluated using stellar wind and stellar atmosphere models (see \citet{crowther2007} for more details). Often errors are not formally propagated into the final values. The stellar-surface magnetic fields of the stars in CWBs is another example of a poorly understood input parameter that is estimated to lie in the range of $10^{2}$ \citep{mathys} to $10^{4}$ G \citep{ignace}. All these effects introduce an ambiguity when disentangling the intrinsic effects of the model from those of model parameters. Further scientific advances in determining the stellar and orbital parameters may improve the situation here. It is expected that hydro-dynamical models that may not need to rely on simplifying assumptions will provide us with a more realistic model of $\gamma$-ray emission from CWBs. To date, the data provided by the \fermi-LAT enabled us to investigate the hypothesis if CWBs constitute a new $\gamma$-ray source population. We show that presently there is no evidence of any population of CWBs that emits detectable $\gamma$-ray emission in the GeV energy regime.\\
However, there is one notable exception: the binary system $\eta$ Car, which is believed to be a CWB consisting of a luminous blue variable (LBV) and an O or a WR star. \fermi-LAT detected a bright point source in spatial coincidence with $\eta$ Car \citep{etacar}. Recently, studies by \citet{klaus} have presented evidence of $\gamma$-ray flux variability that seems to be connected with the orbital period of the $\eta$ Car system. If indeed $\eta$ Car can be confirmed as the first CWB with detectable $\gamma$ ray emission above \fermi-LAT instrumental sensitivity, e.g. by following the source for more than one orbit and finding a resemblance of the modulation pattern as seen from the beginning of the \fermi-LAT observations, there is still the question why we fail to detect other CWBs.\\
The energy carried by the produced $\gamma$-rays is ultimately related to the kinetic power of the stellar wind. Approximately, the magnitude of the $\gamma$-ray emission is proportional to the kinetic power of the stellar wind. The total kinetic power $P_{\mathrm{kin}}$ in the stellar winds of $\eta$ Car can be estimated using the mass loss rates $\dot M$ and terminal velocities $v_{\infty}$ of the constituent stars. Using $\dot M_{LBV}$ = $2.5 \times 10^{-4} \, \mathrm{M}_{\odot} \, \mathrm{yr}^{-1}$ \citep{etamass} and $\dot M_{B}$ = $1.5 \times 10^{-5} \, \mathrm{M}_{\odot} \, \mathrm{yr}^{-1}$ \citep{parkin}, as well as $v_{\infty}^{LBV}$ = $500 \, \mathrm{km} \, \mathrm{s}^{-1}$ \cite{hillier} and $v_{\infty}^{B}$ = $3000 \, \mathrm{km} \mathrm{s}^{-1}$ \citep{parkin}, we arrived at a value of $P_{\mathrm{kin}}\sim 6.2 \times 10^{37} \, \mathrm{erg} \, \mathrm{s}^{-1}$. In the case of WR 140, a system with rather similar orbital parameters, the total kinetic power is $P_{\mathrm{kin}}\approx 1.4 \times 10^{38} \, \mathrm{erg} \, \mathrm{s}^{-1}$, which is 2.2 times higher. Accounting for the fact that $\eta$ Car is $\sim 1.4$ times as distant \citep{monnier} as WR 140 and the detected flux ($\sim 1.5 \times 10^{-7} \, \mathrm{ph} \, \mathrm{cm}^{-2} \, \mathrm{s}^{-1})$ is an order of magnitude larger than the flux upper limits presented in this work, the case can be made that the process by which the kinetic power of the stellar wind is converted to accelerated particles and subsequently to $\gamma$-rays has to be at least an order of magnitude more efficient in the $\eta$ Car system. This simple assessment indicates that the $\eta$ Car system is certainly not a typical representative for the class of CWBs and the processes by which $\gamma$-rays are produced in this unique system will need to be investigated specifically.

\section{Acknowledgements}
\begin{acknowledgements}
The \textit{Fermi} LAT Collaboration acknowledges generous ongoing support
from a number of agencies and institutes that have supported both the
development and the operation of the LAT, as well as scientific data analysis.
These include the National Aeronautics and Space Administration and the
Department of Energy in the United States, the Commissariat \`a l'Energie Atomique
and the Centre National de la Recherche Scientifique / Institut National de Physique
Nucl\'eaire et de Physique des Particules in France, the Agenzia Spaziale Italiana
and the Istituto Nazionale di Fisica Nucleare in Italy, the Ministry of Education,
Culture, Sports, Science and Technology (MEXT), High Energy Accelerator Research
Organization (KEK) and Japan Aerospace Exploration Agency (JAXA) in Japan, and
the K.~A.~Wallenberg Foundation, the Swedish Research Council and the
Swedish National Space Board in Sweden. Additional support for science analysis during the operations phase is gratefully acknowledged from the Istituto Nazionale di Astrofisica in Italy and the Centre National d'\'Etudes Spatiales in France. This research has made use of NASA's Astrophysics Data System. The work presented in this paper was supported by the Austrian Science Fund FWF.

\end{acknowledgements}



\bibliography{References}

\begin{thebibliography}{58}
\expandafter\ifx\csname natexlab\endcsname\relax\def\natexlab#1{#1}\fi

\bibitem[{{Abbott} {et~al.}(1986){Abbott}, {Beiging}, {Churchwell}, \&
  {Torres}}]{Abbott}
{Abbott}, D.~C., {Beiging}, J.~H., {Churchwell}, E., \& {Torres}, A.~V. 1986,
  \apj, 303, 239

\bibitem[{{Abdo} {et~al.}(2010{\natexlab{a}}){Abdo}, {Ackermann}, {Ajello},
  {Allafort}, {Baldini}, {Ballet}, {Barbiellini}, {Baring}, {Bastieri},
  {Bellazzini}, {Blandford}, {Bloom}, {Bonamente}, {Borgland}, {Bouvier},
  {Bregeon}, {Brigida}, {Bruel}, {Burnett}, {Caliandro}, {Cameron}, {Caraveo},
  {Cecchi}, {{\c C}elik}, {Chaty}, {Chekhtman}, {Cheung}, {Chiang}, {Ciprini},
  {Claus}, {Conrad}, {den Hartog}, {Dermer}, {de Angelis}, {de Palma}, {Dib},
  {Dormody}, {Silva}, {Drell}, {Dubois}, {Dumora}, {Enoto}, {Favuzzi},
  {Frailis}, {Fusco}, {Gargano}, {Gehrels}, {Giglietto}, {Giommi}, {Giordano},
  {Giroletti}, {Glanzman}, {Godfrey}, {Grenier}, {Grondin}, {Guiriec},
  {Hadasch}, {Hanabata}, {Harding}, {Hays}, {Israel}, {J{\'o}hannesson},
  {Johnson}, {Kaspi}, {Katagiri}, {Kataoka}, {Kn{\"o}dlseder}, {Kuss}, {Lande},
  {Lee}, {Lemoine-Goumard}, {Longo}, {Loparco}, {Lovellette}, {Lubrano},
  {Makeev}, {Marelli}, {Mazziotta}, {McEnery}, {Mehault}, {Michelson},
  {Mizuno}, {Moiseev}, {Monte}, {Monzani}, {Morselli}, {Moskalenko}, {Murgia},
  {Naumann-Godo}, {Nolan}, {Nuss}, {Ohsugi}, {Okumura}, {Omodei}, {Orlando},
  {Ormes}, {Ozaki}, {Paneque}, {Parent}, {Pepe}, {Pesce-Rollins}, {Piron},
  {Porter}, {Rain{\`o}}, {Rando}, {Razzano}, {Rea}, {Reimer}, {Reimer},
  {Reposeur}, {Ritz}, {Sadrozinski}, {Saz Parkinson}, {Sgr{\`o}}, {Siskind},
  {Smith}, {Spandre}, {Spinelli}, {Strickman}, {Takahashi}, {Tanaka}, {Thayer},
  {Thompson}, {Tibaldo}, {Torres}, {Tosti}, {Tramacere}, {Troja}, {Uchiyama},
  {Usher}, {Vandenbroucke}, {Vasileiou}, {Vianello}, {Vitale}, {Waite},
  {Winer}, {Wood}, {Yang}, \& {Ziegler}}]{magnetars}
{Abdo}, A.~A., {Ackermann}, M., {Ajello}, M., {et~al.} 2010{\natexlab{a}},
  \apjl, 725, L73

\bibitem[{{Abdo} {et~al.}(2010{\natexlab{b}}){Abdo}, {Ackermann}, {Ajello},
  {Allafort}, {Baldini}, {Ballet}, {Barbiellini}, {Bastieri}, {Bechtol},
  {Bellazzini}, {Berenji}, {Blandford}, {Bonamente}, {Borgland}, {Bouvier},
  {Brandt}, {Bregeon}, {Brez}, {Brigida}, {Bruel}, {Buehler}, {Burnett},
  {Caliandro}, {Cameron}, {Caraveo}, {Carrigan}, {Casandjian}, {Cecchi}, {{\c
  C}elik}, {Chaty}, {Chekhtman}, {Cheung}, {Chiang}, {Ciprini}, {Claus},
  {Cohen-Tanugi}, {Cominsky}, {Conrad}, {Dermer}, {de Palma}, {Digel}, {Silva},
  {Drell}, {Dubois}, {Dumora}, {Favuzzi}, {Fegan}, {Ferrara}, {Frailis},
  {Fukazawa}, {Fusco}, {Gargano}, {Gehrels}, {Germani}, {Giglietto},
  {Giordano}, {Godfrey}, {Grenier}, {Grondin}, {Grove}, {Guillemot}, {Guiriec},
  {Hadasch}, {Hanabata}, {Harding}, {Hayashida}, {Hays}, {Hill}, {Horan},
  {Hughes}, {Itoh}, {Jackson}, {J{\'o}hannesson}, {Johnson}, {Johnson},
  {Kamae}, {Katagiri}, {Kataoka}, {Kerr}, {Kn{\"o}dlseder}, {Kuss}, {Lande},
  {Latronico}, {Lee}, {Lemoine-Goumard}, {Livingstone}, {Llena Garde}, {Longo},
  {Loparco}, {Lovellette}, {Lubrano}, {Makeev}, {Mazziotta}, {McEnery},
  {Mehault}, {Michelson}, {Mitthumsiri}, {Mizuno}, {Moiseev}, {Monte},
  {Monzani}, {Morselli}, {Moskalenko}, {Murgia}, {Nakamori}, {Naumann-Godo},
  {Nolan}, {Norris}, {Nuss}, {Ohsugi}, {Okumura}, {Omodei}, {Orlando}, {Ormes},
  {Ozaki}, {Panetta}, {Parent}, {Pelassa}, {Pepe}, {Pesce-Rollins}, {Piron},
  {Porter}, {Rain{\`o}}, {Rando}, {Razzano}, {Reimer}, {Reimer}, {Reposeur},
  {Rodriguez}, {Romani}, {Roth}, {Sadrozinski}, {Sander}, {Saz Parkinson},
  {Scargle}, {Sgr{\`o}}, {Siskind}, {Smith}, {Smith}, {Spandre}, {Spinelli},
  {Strickman}, {Suson}, {Takahashi}, {Takahashi}, {Tanaka}, {Thayer}, {Thayer},
  {Thompson}, {Tibaldo}, {Tibolla}, {Torres}, {Tosti}, {Tramacere}, {Uchiyama},
  {Usher}, {Vandenbroucke}, {Vasileiou}, {Vilchez}, {Vitale}, {Waite},
  {Wallace}, {Wang}, {Winer}, {Wood}, {Yang}, {Ylinen}, \&
  {Ziegler}}]{Abdo2010}
{Abdo}, A.~A., {Ackermann}, M., {Ajello}, M., {et~al.} 2010{\natexlab{b}},
  \apj, 723, 649

\bibitem[{{Abdo} {et~al.}(2010{\natexlab{c}}){Abdo}, {Ackermann}, {Ajello},
  {Allafort}, {Baldini}, {Ballet}, {Barbiellini}, {Bastieri}, {Bechtol},
  {Bellazzini}, {Berenji}, {Blandford}, {Bonamente}, {Borgland}, {Bouvier},
  {Brandt}, {Bregeon}, {Brez}, {Brigida}, {Bruel}, {Buehler}, {Burnett},
  {Caliandro}, {Cameron}, {Caraveo}, {Carrigan}, {Casandjian}, {Cecchi}, {{\c
  C}elik}, {Chaty}, {Chekhtman}, {Cheung}, {Chiang}, {Ciprini}, {Claus},
  {Cohen-Tanugi}, {Cominsky}, {Conrad}, {Dermer}, {de Palma}, {Digel}, {Silva},
  {Drell}, {Dubois}, {Dumora}, {Favuzzi}, {Fegan}, {Ferrara}, {Frailis},
  {Fukazawa}, {Fusco}, {Gargano}, {Gehrels}, {Germani}, {Giglietto},
  {Giordano}, {Godfrey}, {Grenier}, {Grondin}, {Grove}, {Guillemot}, {Guiriec},
  {Hadasch}, {Hanabata}, {Harding}, {Hayashida}, {Hays}, {Hill}, {Horan},
  {Hughes}, {Itoh}, {Jackson}, {J{\'o}hannesson}, {Johnson}, {Johnson},
  {Kamae}, {Katagiri}, {Kataoka}, {Kerr}, {Kn{\"o}dlseder}, {Kuss}, {Lande},
  {Latronico}, {Lee}, {Lemoine-Goumard}, {Livingstone}, {Llena Garde}, {Longo},
  {Loparco}, {Lovellette}, {Lubrano}, {Makeev}, {Mazziotta}, {McEnery},
  {Mehault}, {Michelson}, {Mitthumsiri}, {Mizuno}, {Moiseev}, {Monte},
  {Monzani}, {Morselli}, {Moskalenko}, {Murgia}, {Nakamori}, {Naumann-Godo},
  {Nolan}, {Norris}, {Nuss}, {Ohsugi}, {Okumura}, {Omodei}, {Orlando}, {Ormes},
  {Ozaki}, {Panetta}, {Parent}, {Pelassa}, {Pepe}, {Pesce-Rollins}, {Piron},
  {Porter}, {Rain{\`o}}, {Rando}, {Razzano}, {Reimer}, {Reimer}, {Reposeur},
  {Rodriguez}, {Romani}, {Roth}, {Sadrozinski}, {Sander}, {Saz Parkinson},
  {Scargle}, {Sgr{\`o}}, {Siskind}, {Smith}, {Smith}, {Spandre}, {Spinelli},
  {Strickman}, {Suson}, {Takahashi}, {Takahashi}, {Tanaka}, {Thayer}, {Thayer},
  {Thompson}, {Tibaldo}, {Tibolla}, {Torres}, {Tosti}, {Tramacere}, {Uchiyama},
  {Usher}, {Vandenbroucke}, {Vasileiou}, {Vilchez}, {Vitale}, {Waite},
  {Wallace}, {Wang}, {Winer}, {Wood}, {Yang}, {Ylinen}, \& {Ziegler}}]{etacar}
{Abdo}, A.~A., {Ackermann}, M., {Ajello}, M., {et~al.} 2010{\natexlab{c}},
  \apj, 723, 649

\bibitem[{{Ackermann} {et~al.}(2012{\natexlab{a}}){Ackermann}, {Ajello},
  {Allafort}, {Baldini}, {Ballet}, {Barbiellini}, {Bastieri}, {Belfiore},
  {Bellazzini}, {Berenji}, {Blandford}, {Bloom}, {Bonamente}, {Borgland},
  {Bottacini}, {Bregeon}, {Brigida}, {Bruel}, {Buehler}, {Buson}, {Caliandro},
  {Cameron}, {Caraveo}, {Casandjian}, {Cecchi}, {Chekhtman}, {Ciprini},
  {Claus}, {Cohen-Tanugi}, {de Angelis}, {de Palma}, {Dermer}, {Silva},
  {Drell}, {Dumora}, {Favuzzi}, {Fegan}, {Focke}, {Fortin}, {Fukazawa},
  {Fusco}, {Gargano}, {Germani}, {Giglietto}, {Giordano}, {Giroletti},
  {Glanzman}, {Godfrey}, {Grenier}, {Guillemot}, {Guiriec}, {Hadasch},
  {Hanabata}, {Harding}, {Hayashida}, {Hayashi}, {Hays}, {J{\'o}hannesson},
  {Johnson}, {Kamae}, {Katagiri}, {Kataoka}, {Kerr}, {Kn{\"o}dlseder}, {Kuss},
  {Lande}, {Latronico}, {Lee}, {Longo}, {Loparco}, {Lott}, {Lovellette},
  {Lubrano}, {Martin}, {Mazziotta}, {McEnery}, {Mehault}, {Michelson},
  {Mitthumsiri}, {Mizuno}, {Monte}, {Monzani}, {Morselli}, {Moskalenko},
  {Murgia}, {Naumann-Godo}, {Nolan}, {Norris}, {Nuss}, {Ohsugi}, {Okumura},
  {Omodei}, {Orlando}, {Ormes}, {Ozaki}, {Paneque}, {Parent}, {Pesce-Rollins},
  {Pierbattista}, {Piron}, {Porter}, {Rain{\`o}}, {Rando}, {Razzano}, {Reimer},
  {Reposeur}, {Ritz}, {Saz Parkinson}, {Sgr{\`o}}, {Siskind}, {Smith},
  {Spinelli}, {Strong}, {Takahashi}, {Tanaka}, {Thayer}, {Thayer}, {Thompson},
  {Tibaldo}, {Torres}, {Tosti}, {Tramacere}, {Troja}, {Uchiyama},
  {Vandenbroucke}, {Vasileiou}, {Vianello}, {Vitale}, {Waite}, {Wang}, {Winer},
  {Wood}, {Yang}, {Zimmer}, \& {Bontemps}}]{luigi_cygnus}
{Ackermann}, M., {Ajello}, M., {Allafort}, A., {et~al.} 2012{\natexlab{a}},
  \aap, 538, A71

\bibitem[{{Ackermann} {et~al.}(2012{\natexlab{b}}){Ackermann}, {Ajello},
  {Atwood}, {Baldini}, {Ballet}, {Barbiellini}, {Bastieri}, {Bechtol},
  {Bellazzini}, {Berenji}, {Blandford}, {Bloom}, {Bonamente}, {Borgland},
  {Brandt}, {Bregeon}, {Brigida}, {Bruel}, {Buehler}, {Buson}, {Caliandro},
  {Cameron}, {Caraveo}, {Cavazzuti}, {Cecchi}, {Charles}, {Chekhtman},
  {Chiang}, {Ciprini}, {Claus}, {Cohen-Tanugi}, {Conrad}, {Cutini}, {de
  Angelis}, {de Palma}, {Dermer}, {Digel}, {Silva}, {Drell}, {Drlica-Wagner},
  {Falletti}, {Favuzzi}, {Fegan}, {Ferrara}, {Focke}, {Fortin}, {Fukazawa},
  {Funk}, {Fusco}, {Gaggero}, {Gargano}, {Germani}, {Giglietto}, {Giordano},
  {Giroletti}, {Glanzman}, {Godfrey}, {Grove}, {Guiriec}, {Gustafsson},
  {Hadasch}, {Hanabata}, {Harding}, {Hayashida}, {Hays}, {Horan}, {Hou},
  {Hughes}, {J{\'o}hannesson}, {Johnson}, {Johnson}, {Kamae}, {Katagiri},
  {Kataoka}, {Kn{\"o}dlseder}, {Kuss}, {Lande}, {Latronico}, {Lee},
  {Lemoine-Goumard}, {Longo}, {Loparco}, {Lott}, {Lovellette}, {Lubrano},
  {Mazziotta}, {McEnery}, {Michelson}, {Mitthumsiri}, {Mizuno}, {Monte},
  {Monzani}, {Morselli}, {Moskalenko}, {Murgia}, {Naumann-Godo}, {Norris},
  {Nuss}, {Ohsugi}, {Okumura}, {Omodei}, {Orlando}, {Ormes}, {Paneque},
  {Panetta}, {Parent}, {Pesce-Rollins}, {Pierbattista}, {Piron}, {Pivato},
  {Porter}, {Rain{\`o}}, {Rando}, {Razzano}, {Razzaque}, {Reimer}, {Reimer},
  {Sadrozinski}, {Sgr{\`o}}, {Siskind}, {Spandre}, {Spinelli}, {Strong},
  {Suson}, {Takahashi}, {Tanaka}, {Thayer}, {Thayer}, {Thompson}, {Tibaldo},
  {Tinivella}, {Torres}, {Tosti}, {Troja}, {Usher}, {Vandenbroucke},
  {Vasileiou}, {Vianello}, {Vitale}, {Waite}, {Wang}, {Winer}, {Wood}, {Wood},
  {Yang}, {Ziegler}, \& {Zimmer}}]{strong_diffuse2}
{Ackermann}, M., {Ajello}, M., {Atwood}, W.~B., {et~al.} 2012{\natexlab{b}},
  \apj, 750, 3

\bibitem[{{Ackermann} {et~al.}(2012{\natexlab{c}})}]{pass7}
{Ackermann}, M. {et~al.} 2012{\natexlab{c}}, \apjs, 203, 4

\bibitem[{{Aliu} {et~al.}(2008){Aliu}, {Anderhub}, {Antonelli}, {Antoranz},
  {Backes}, {Baixeras}, {Barrio}, {Bartko}, {Bastieri}, {Becker}, {Bednarek},
  {Berger}, {Bernardini}, {Bigongiari}, {Biland}, {Bock}, {Bonnoli}, {Bordas},
  {Bosch-Ramon}, {Bretz}, {Britvitch}, {Camara}, {Carmona}, {Chilingarian},
  {Commichau}, {Contreras}, {Cortina}, {Costado}, {Covino}, {Curtef}, {Dazzi},
  {De Angelis}, {De Cea del Pozo}, {de los Reyes}, {De Lotto}, {De Maria}, {De
  Sabata}, {Delgado Mendez}, {Dominguez}, {Dorner}, {Doro}, {Errando},
  {Fagiolini}, {Ferenc}, {Fern{\'a}ndez}, {Firpo}, {Fonseca}, {Font},
  {Galante}, {Garc{\'{\i}}a L{\'o}pez}, {Garczarczyk}, {Gaug}, {Goebel},
  {Hayashida}, {Herrero}, {H{\"o}hne}, {Hose}, {Hsu}, {Huber}, {Jogler},
  {Kranich}, {La Barbera}, {Laille}, {Leonardo}, {Lindfors}, {Lombardi},
  {Longo}, {L{\'o}pez}, {Lorenz}, {Majumdar}, {Maneva}, {Mankuzhiyil},
  {Mannheim}, {Maraschi}, {Mariotti}, {Mart{\'{\i}}nez}, {Mazin}, {Meucci},
  {Meyer}, {Miranda}, {Mirzoyan}, {Moles}, {Moralejo}, {Nieto}, {Nilsson},
  {Ninkovic}, {O{\~n}a-Wilhelmi}, {Otte}, {Oya}, {Paoletti}, {Paredes},
  {Pasanen}, {Pascoli}, {Pauss}, {Pegna}, {Perez-Torres}, {Persic}, {Peruzzo},
  {Piccioli}, {Prada}, {Prandini}, {Puchades}, {Raymers}, {Rhode}, {Rib{\'o}},
  {Rico}, {Rissi}, {Robert}, {R{\"u}gamer}, {Saggion}, {Saito}, {Salvati},
  {Sanchez-Conde}, {Sartori}, {Satalecka}, {Scalzotto}, {Scapin}, {Schweizer},
  {Shayduk}, {Shinozaki}, {Shore}, {Sidro}, {Sierpowska-Bartosik},
  {Sillanp{\"a}{\"a}}, {Sobczynska}, {Spanier}, {Stamerra}, {Stark}, {Takalo},
  {Tavecchio}, {Temnikov}, {Tescaro}, {Teshima}, {Tluczykont}, {Torres},
  {Turini}, {Vankov}, {Venturini}, {Vitale}, {Wagner}, {Wittek}, {Zabalza},
  {Zandanel}, {Zanin}, \& {Zapatero}}]{Alui}
{Aliu}, E., {Anderhub}, H., {Antonelli}, L.~A., {et~al.} 2008, \apjl, 685, L71

\bibitem[{{Atwood} {et~al.}(2009){Atwood}, {Abdo}, {Ackermann}, {Althouse},
  {Anderson}, {Axelsson}, {Baldini}, {Ballet}, {Band}, {Barbiellini}, \&
  et~al.}]{lat}
{Atwood}, W.~B., {Abdo}, A.~A., {Ackermann}, M., {et~al.} 2009, \apj, 697, 1071

\bibitem[{{Benaglia} \& {Romero}(2003)}]{Romero2003}
{Benaglia}, P. \& {Romero}, G.~E. 2003, \aap, 399, 1121

\bibitem[{{Chapman} {et~al.}(1999){Chapman}, {Leitherer}, {Koribalski},
  {Bouter}, \& {Storey}}]{Chapman}
{Chapman}, J.~M., {Leitherer}, C., {Koribalski}, B., {Bouter}, R., \& {Storey},
  M. 1999, \apj, 518, 890

\bibitem[{{Contreras} \& {Rodr{\'{\i}}guez}(1999)}]{Contreras}
{Contreras}, M.~E. \& {Rodr{\'{\i}}guez}, L.~F. 1999, \apj, 515, 762

\bibitem[{{Crowther}(1997)}]{crowtherTemp}
{Crowther}, P.~A. 1997, in IAU Symposium, Vol. 189, The effective temperatures
  of hot stars., ed. T.~R. {Bedding}, A.~J. {Booth}, \& J.~{Davis}, 137--146

\bibitem[{{Crowther}(2007)}]{crowther2007}
{Crowther}, P.~A. 2007, \araa, 45, 177

\bibitem[{{De Marco} {et~al.}(2000){De Marco}, {Schmutz}, {Crowther},
  {Hillier}, {Dessart}, {de Koter}, \& {Schweickhardt}}]{DeMarco2}
{De Marco}, O., {Schmutz}, W., {Crowther}, P.~A., {et~al.} 2000, \aap, 358, 187

\bibitem[{{Dougherty} {et~al.}(2005){Dougherty}, {Beasley}, {Claussen},
  {Zauderer}, \& {Bolingbroke}}]{dougherty2}
{Dougherty}, S.~M., {Beasley}, A.~J., {Claussen}, M.~J., {Zauderer}, B.~A., \&
  {Bolingbroke}, N.~J. 2005, \apj, 623, 447

\bibitem[{{Dougherty} {et~al.}(2003){Dougherty}, {Pittard}, {Kasian}, {Coker},
  {Williams}, \& {Lloyd}}]{dougherty3}
{Dougherty}, S.~M., {Pittard}, J.~M., {Kasian}, L., {et~al.} 2003, \aap, 409,
  217

\bibitem[{{Dougherty} {et~al.}(2000){Dougherty}, {Williams}, \&
  {Pollacco}}]{doughertyWR146}
{Dougherty}, S.~M., {Williams}, P.~M., \& {Pollacco}, D.~L. 2000, \mnras, 316,
  143

\bibitem[{{Dougherty} {et~al.}(1996){Dougherty}, {Williams}, {van der Hucht},
  {Bode}, \& {Davis}}]{dougherty1}
{Dougherty}, S.~M., {Williams}, P.~M., {van der Hucht}, K.~A., {Bode}, M.~F.,
  \& {Davis}, R.~J. 1996, \mnras, 280, 963

\bibitem[{{Dubus} {et~al.}(2008){Dubus}, {Cerutti}, \& {Henri}}]{cerutti}
{Dubus}, G., {Cerutti}, B., \& {Henri}, G. 2008, \aap, 477, 691

\bibitem[{{Eichler} \& {Usov}(1993)}]{eichler1993}
{Eichler}, D. \& {Usov}, V. 1993, \apj, 402, 271

\bibitem[{{Feldman} \& {Cousins}(1998)}]{fc}
{Feldman}, G.~J. \& {Cousins}, R.~D. 1998, \prd, 57, 3873

\bibitem[{{Hillier} {et~al.}(2001){Hillier}, {Davidson}, {Ishibashi}, \&
  {Gull}}]{hillier}
{Hillier}, D.~J., {Davidson}, K., {Ishibashi}, K., \& {Gull}, T. 2001, \apj,
  553, 837

\bibitem[{{Ignace} {et~al.}(1998){Ignace}, {Cassinelli}, \&
  {Bjorkman}}]{ignace}
{Ignace}, R., {Cassinelli}, J.~P., \& {Bjorkman}, J.~E. 1998, \apj, 505, 910

\bibitem[{{Lef{\`e}vre} {et~al.}(2005){Lef{\`e}vre}, {Marchenko}, {L{\'e}pine},
  {Moffat}, {Acker}, {Harries}, {Annuk}, {Bohlender}, {Demers}, {Grosdidier},
  {Hill}, {Morrison}, {Knauth}, {Skalkowski}, \& {Viti}}]{wr137orbit}
{Lef{\`e}vre}, L., {Marchenko}, S.~V., {L{\'e}pine}, S., {et~al.} 2005, \mnras,
  360, 141

\bibitem[{{Marchenko} {et~al.}(2003){Marchenko}, {Moffat}, {Ballereau},
  {Chauville}, {Zorec}, {Hill}, {Annuk}, {Corral}, {Demers}, {Eenens}, {Panov},
  {Seggewiss}, {Thomson}, \& {Villar-Sbaffi}}]{Marchenko}
{Marchenko}, S.~V., {Moffat}, A.~F.~J., {Ballereau}, D., {et~al.} 2003, \apj,
  596, 1295

\bibitem[{{Mathys}(1999)}]{mathys}
{Mathys}, G. 1999, in Lecture Notes in Physics, Berlin Springer Verlag, Vol.
  523, IAU Colloq. 169: Variable and Non-spherical Stellar Winds in Luminous
  Hot Stars, ed. {B.~Wolf, O.~Stahl, \& A.~W.~Fullerton}, 95

\bibitem[{{Mattox} {et~al.}(1996){Mattox}, {Bertsch}, {Chiang}, {Dingus},
  {Digel}, {Esposito}, {Fierro}, {Hartman}, {Hunter}, {Kanbach}, {Kniffen},
  {Lin}, {Macomb}, {Mayer-Hasselwander}, {Michelson}, {von Montigny},
  {Mukherjee}, {Nolan}, {Ramanamurthy}, {Schneid}, {Sreekumar}, {Thompson}, \&
  {Willis}}]{mattox}
{Mattox}, J.~R., {Bertsch}, D.~L., {Chiang}, J., {et~al.} 1996, \apj, 461, 396

\bibitem[{{Millour} {et~al.}(2007){Millour}, {Petrov}, {Chesneau}, {Bonneau},
  {Dessart}, {Bechet}, {Tallon-Bosc}, {Tallon}, {Thi{\'e}baut}, {Vakili},
  {Malbet}, {Mourard}, {Antonelli}, {Beckmann}, {Bresson}, {Chelli},
  {Dugu{\'e}}, {Duvert}, {Gennari}, {Gl{\"u}ck}, {Kern}, {Lagarde}, {Le
  Coarer}, {Lisi}, {Perraut}, {Puget}, {Rantakyr{\"o}}, {Robbe-Dubois},
  {Roussel}, {Tatulli}, {Weigelt}, {Zins}, {Accardo}, {Acke}, {Agabi},
  {Altariba}, {Arezki}, {Aristidi}, {Baffa}, {Behrend}, {Bl{\"o}cker},
  {Bonhomme}, {Busoni}, {Cassaing}, {Clausse}, {Colin}, {Connot},
  {Delboulb{\'e}}, {Domiciano de Souza}, {Driebe}, {Feautrier}, {Ferruzzi},
  {Forveille}, {Fossat}, {Foy}, {Fraix-Burnet}, {Gallardo}, {Giani}, {Gil},
  {Glentzlin}, {Heiden}, {Heininger}, {Hernandez Utrera}, {Hofmann}, {Kamm},
  {Kiekebusch}, {Kraus}, {Le Contel}, {Le Contel}, {Lesourd}, {Lopez}, {Lopez},
  {Magnard}, {Marconi}, {Mars}, {Martinot-Lagarde}, {Mathias}, {M{\`e}ge},
  {Monin}, {Mouillet}, {Nussbaum}, {Ohnaka}, {Pacheco}, {Perrier}, {Rabbia},
  {Rebattu}, {Reynaud}, {Richichi}, {Robini}, {Sacchettini}, {Schertl},
  {Sch{\"o}ller}, {Solscheid}, {Spang}, {Stee}, {Stefanini}, {Tasso}, {Testi},
  {von der L{\"u}he}, {Valtier}, {Vannier}, \& {Ventura}}]{Millour2007}
{Millour}, F., {Petrov}, R.~G., {Chesneau}, O., {et~al.} 2007, \aap, 464, 107

\bibitem[{{Monnier} {et~al.}(2011){Monnier}, {Zhao}, {Pedretti},
  {Millan-Gabet}, {Berger}, {Traub}, {Schloerb}, {ten Brummelaar}, {McAlister},
  {Ridgway}, {Sturmann}, {Sturmann}, {Turner}, {Baron}, {Kraus}, {Tannirkulam},
  \& {Williams}}]{monnier}
{Monnier}, J.~D., {Zhao}, M., {Pedretti}, E., {et~al.} 2011, \apjl, 742, L1

\bibitem[{{Morris} {et~al.}(2000){Morris}, {van der Hucht}, {Crowther},
  {Hillier}, {Dessart}, {Williams}, \& {Willis}}]{MorrisWR147dist}
{Morris}, P.~W., {van der Hucht}, K.~A., {Crowther}, P.~A., {et~al.} 2000,
  \aap, 353, 624

\bibitem[{{Nolan} {et~al.}(2012){Nolan}, {Abdo}, {Ackermann}, {Ajello},
  {Allafort}, {Antolini}, {Atwood}, {Axelsson}, {Baldini}, {Ballet}, \&
  et~al.}]{2fgl}
{Nolan}, P.~L., {Abdo}, A.~A., {Ackermann}, M., {et~al.} 2012, \apjs, 199, 31

\bibitem[{{North} {et~al.}(2007){North}, {Tuthill}, {Tango}, \&
  {Davis}}]{NorthWR}
{North}, J.~R., {Tuthill}, P.~G., {Tango}, W.~J., \& {Davis}, J. 2007, \mnras,
  377, 415

\bibitem[{{O'Connor} {et~al.}(2005){O'Connor}, {Dougherty}, {Pittard}, \&
  {Williams}}]{Connor}
{O'Connor}, E.~P., {Dougherty}, S.~M., {Pittard}, J.~M., \& {Williams}, P.~M.
  2005, in Massive Stars and High-Energy Emission in OB Associations, ed.
  {G.~Rauw, Y.~Naz{\'e}, R.~Blomme, \& E.~Gosset}, 81--84

\bibitem[{{Parkin} {et~al.}(2009){Parkin}, {Pittard}, {Corcoran}, {Hamaguchi},
  \& {Stevens}}]{parkin}
{Parkin}, E.~R., {Pittard}, J.~M., {Corcoran}, M.~F., {Hamaguchi}, K., \&
  {Stevens}, I.~R. 2009, \mnras, 394, 1758

\bibitem[{{Pittard} \& {Corcoran}(2002)}]{etamass}
{Pittard}, J.~M. \& {Corcoran}, M.~F. 2002, \aap, 383, 636

\bibitem[{{Pittard} \& {Dougherty}(2006)}]{Pittard2006}
{Pittard}, J.~M. \& {Dougherty}, S.~M. 2006, \mnras, 372, 801

\bibitem[{{Pollock}(1987)}]{pollock1987}
{Pollock}, A.~M.~T. 1987, \aap, 171, 135

\bibitem[{{Reimer} {et~al.}(2006){Reimer}, {Pohl}, \& {Reimer}}]{Reimer2006}
{Reimer}, A., {Pohl}, M., \& {Reimer}, O. 2006, \apj, 644, 1118

\bibitem[{{Reimer} \& {Reimer}(2007)}]{Reimer2007}
{Reimer}, A. \& {Reimer}, O. 2007, in American Institute of Physics Conference
  Series, Vol. 921, The First GLAST Symposium, ed. {S.~Ritz, P.~Michelson, \&
  C.~A.~Meegan}, 217--219

\bibitem[{{Reimer} \& {Reimer}(2009{\natexlab{a}})}]{ReimerAIP2009}
{Reimer}, A. \& {Reimer}, O. 2009{\natexlab{a}}, in AIP Conf. Proc, Vol. 1112,
  Massive stars in colliding wind systems: the high-energy gamma-ray
  perspective, ed. D.~{Bastieri} \& R.~{Rando}, 43--53

\bibitem[{{Reimer} \& {Reimer}(2009{\natexlab{b}})}]{Reimer2009}
{Reimer}, A. \& {Reimer}, O. 2009{\natexlab{b}}, \apj, 694, 1139

\bibitem[{{Reitberger} {et~al.}(2012){Reitberger}, {Reimer}, {Reimer},
  {Werner}, {Egberts}, \& {Takahashi}}]{klaus}
{Reitberger}, K., {Reimer}, O., {Reimer}, A., {et~al.} 2012, \aap, 544, A98

\bibitem[{{Romero} {et~al.}(1999){Romero}, {Benaglia}, \&
  {Torres}}]{Romero1999}
{Romero}, G.~E., {Benaglia}, P., \& {Torres}, D.~F. 1999, \aap, 348, 868

\bibitem[{{Setia Gunawan} {et~al.}(2001{\natexlab{a}}){Setia Gunawan}, {de
  Bruyn}, {van der Hucht}, \& {Williams}}]{SetiaWR147period}
{Setia Gunawan}, D.~Y.~A., {de Bruyn}, A.~G., {van der Hucht}, K.~A., \&
  {Williams}, P.~M. 2001{\natexlab{a}}, \aap, 368, 484

\bibitem[{{Setia Gunawan} {et~al.}(2001{\natexlab{b}}){Setia Gunawan}, {van der
  Hucht}, {Williams}, {Henrichs}, {Kaper}, {Stickland}, \&
  {Wamsteker}}]{SetiaWR140velocity}
{Setia Gunawan}, D.~Y.~A., {van der Hucht}, K.~A., {Williams}, P.~M., {et~al.}
  2001{\natexlab{b}}, \aap, 376, 460

\bibitem[{{Tavani} {et~al.}(2009)}]{agile}
{Tavani}, M. {et~al.} 2009, \apjl, 698, L142

\bibitem[{{Vacca} {et~al.}(1996){Vacca}, {Garmany}, \& {Shull}}]{vacca}
{Vacca}, W.~D., {Garmany}, C.~D., \& {Shull}, J.~M. 1996, \apj, 460, 914

\bibitem[{{van der Hucht}(2001)}]{vanderhucht}
{van der Hucht}, K.~A. 2001, \nar, 45, 135

\bibitem[{{van der Hucht}(2006)}]{catannex}
{van der Hucht}, K.~A. 2006, \aap, 458, 453

\bibitem[{{van der Hucht} {et~al.}(1997){van der Hucht}, {Schrijver},
  {Stenholm}, {Lundstrom}, {Moffat}, {Marchenko}, {Seggewiss}, {Setia Gunawan},
  {Sutantyo}, {van den Heuvel}, {de Cuyper}, \& {Gomez}}]{huchtwr11}
{van der Hucht}, K.~A., {Schrijver}, H., {Stenholm}, B., {et~al.} 1997, \na, 2,
  245

\bibitem[{{van der Hucht} {et~al.}(2002){van der Hucht}, {Setia Gunawan},
  {Williams}, {de Bruyn}, \& {Spoelstra}}]{vanHuchtwr146dist}
{van der Hucht}, K.~A., {Setia Gunawan}, D.~Y.~A., {Williams}, P.~M., {de
  Bruyn}, A.~G., \& {Spoelstra}, T.~A.~T. 2002, in Astronomical Society of the
  Pacific Conference Series, Vol. 260, Interacting Winds from Massive Stars,
  ed. {A.~F.~J.~Moffat \& N.~St-Louis}, 297

\bibitem[{{Wallace} {et~al.}(2001){Wallace}, {Nelan}, {Leitherer}, {Gies},
  {Moffat}, \& {Shara}}]{Wallace}
{Wallace}, D.~J., {Nelan}, E., {Leitherer}, C., {et~al.} 2001, in Astrophysics
  and Space Science Library, Vol. 264, The Influence of Binaries on Stellar
  Population Studies, ed. {D.~Vanbeveren}, 173

\bibitem[{{White} \& {Becker}(1995)}]{white1995}
{White}, R.~L. \& {Becker}, R.~H. 1995, \apj, 451, 352

\bibitem[{{Williams} {et~al.}(1997){Williams}, {Dougherty}, {Davis}, {van der
  Hucht}, {Bode}, \& {Setia Gunawan}}]{Williams}
{Williams}, P.~M., {Dougherty}, S.~M., {Davis}, R.~J., {et~al.} 1997, \mnras,
  289, 10

\bibitem[{{Williams} {et~al.}(1990){Williams}, {van der Hucht}, {Pollock},
  {Florkowski}, {van der Woerd}, \& {Wamsteker}}]{Williams1990}
{Williams}, P.~M., {van der Hucht}, K.~A., {Pollock}, A.~M.~T., {et~al.} 1990,
  \mnras, 243, 662

\bibitem[{{Willis} {et~al.}(1997){Willis}, {Dessart}, {Crowther}, {Morris},
  {Maeder}, {Conti}, \& {van der Hucht}}]{Willis}
{Willis}, A.~J., {Dessart}, L., {Crowther}, P.~A., {et~al.} 1997, \mnras, 290,
  371

\bibitem[{{Zhekov} \& {Park}(2010)}]{Zhekov2}
{Zhekov}, S.~A. \& {Park}, S. 2010, \apj, 721, 518

\end{thebibliography}

\clearpage
\onecolumn


\clearpage

\appendix
\section{Parameters characterizing the analysed CWB sample}

\begin{table*}[h]
\caption{Parameters characterizing the analysed CWB sample}
\label{wrparam}
\centering
\resizebox{1.0\textwidth}{!}
{
\renewcommand{\arraystretch}{1.5}
\begin{tabular}{l c c c c c c c c}
\hline\hline
Parameter		&Unit 								&WR 11					&WR 70			&WR 125					&WR 137					&WR 140					&WR 146		&WR 147\\
\hline
Sp. Type		&									& WC8 + O7.5 (4)		& WC9vd + B0I 	&WC7ed + O9III 			&WC7pd + O9				&WC7pd + O4-5			&WC6 + O8	&WN8 + B0.5V\\
$l$			&[$^\circ$]									&262.80					&322.34 		&54.44					&74.33					&80.93 					&80.56		&79.85\\
$b$		&[$^\circ$]								&−7.69					&−1.81 			&1.06 					&1.09 					&4.18 					&0.45 		&−0.32\\
$d_{L}$		&[pc]								&$258^{+41}_{-31}$ (1)	&1910			&3060 					&2380 					&1100 					&720		&$650^{+130}_{-110}$ (9)\\
			&									&$368^{+38}_{-13}$ (3) 	&				&2130 (5) 				&1600 (6) 				&1850 (7) 				&1250 (8)	&\\
	&									& 	&				&				& 				&$1670 \pm 0.03$ (21) 				&	&\\
$P$		&[d]								&$78.53 \pm 0.01$ 		& $> 4000$		& $> 6600$				&$4766 \pm 66$ (20)		&$2899 \pm 1.3$ (10)	&$\sim 300$ yr (11) 			& $> 1350$ yr (12) \\
	&									& 	&				&				& 				&$2896.35 \pm 0.20$ (21) 				&	&\\
$i$	&[$^\circ$]							&$65.5 \pm 0.4$ (2)		&(...) 			&(...) 					&(...)					&$122 \pm 5$ (7) 		&(...)			&(..)\\
	&									& 	&				&				& 				&$119.6 \pm 0.5$ (21) 				&	&\\
$e$	&									&$0.334 \pm 0.003$ (2)	&(...) 			&(...)					&$0.178 \pm 0.04$ (20)	&$0.881 \pm 0.005$ (10)	&(...) 			&(...)\\
	&									& 	&				&				& 				&$0.8964^{+0.0004}_{-0.0007}$ (21) 				&	&\\
$M_{WR}$	 	
	&[$\mathrm{M}_{\odot}$]						&$9.5 \pm 1$ (4)		&$< 20$			&(...)					&$4.4 \pm 1.5$ (20)		&$20 \pm 4$	(7)	 		&(...)		&(...)\\
										&									& 	&				&				& 				&$14.9 \pm 0.5$ (21) 				&	&\\
$M_{OB}$		&[$\mathrm{M}_{\odot}$]						&$30 \pm 2$ (4)			&$>5$ 			&(...)					&$20 \pm 2$	(20)		&$54 \pm 10$ (7)		&(...)			&(...)\\
										&									& 	&				&				& 				&$35.9 \pm 1.3$ (21) 				&	&\\
$L_{OB}$	 	&[$10^{5} \, \mathrm{L}_{\odot}$]			&$2.1 \pm 0.3$ (4)		&(...)			&(...)					&(...)					&15.1 (7)				&1.0 (13)			&0.5 (9)\\
					&									&$2.8$ (2)				&				&						&						&						&			&\\
$T_{OB}$		&[K]								&$35000 \pm 300$ (4)	&(...)			&(...)					&(...)					&47400 (17)				&35700 (17)			&28500 (19)\\
$\dot M_{WR}$	&[$10^{-5} \, \mathrm{M}_{\odot} \, \mathrm{y}^{-1}$] &$3$ (2) 				&(...)			&(...)					&3.3 (6)				&4.3 (7) 				&4	(16)	&2.4 (9)\\
	&									& 						&				&						&						&						&2.6 (18)	&\\
$\dot M_{OB}$ &[$10^{-7} \, \mathrm{M}_{\odot} \, \mathrm{y}^{-1}$] &$1.78 \pm 0.37$ (4)	&(...)			&(...)					&(...)					&87						&80 (16)	&4 (12)\\
$v^{WR}_{\infty}$ &[$\mathrm{km} \, \mathrm{s}^{-1}$] 					&$1550$			 (4)	&1150			&2900					&1900					&2860 (14)				&$2700 \pm 200$	(18)	&950 (9)\\
$v^{OB}_{\infty}$	&[$\mathrm{km} \, \mathrm{s}^{-1}$] 					&$2500 \pm 250$  (4)	&(...)			&(...)					&(...)					&3100 (15)				&$1600 \pm 480$ (16)	&800 (12)\\
&									& & & & & & &\\
\end{tabular}
}
\tablefoot{Parameters are Sp. Type: spectral Type; $l$: Galactic longitude; $b$: Galactic latitude; $d_{L}$: distance; $P$: orbital period; $i$: inclination; $e$: eccentricity; $M_{WR}$: mass of the WR star; $M_{OB}$: mass of OB-companion star; $L_{OB}$: OB-star luminosity; $T_{OB}$: surface temperature of OB-star; $\dot M_{WR}$: WR star mass loss rate; $\dot M_{OB}$: OB-star  mass loss rate; $v^{WR}_{\infty}$: terminal velocity of WR star stellar wind; $v^{OB}_{\infty}$: terminal velocity of OB-star stellar wind.}
\tablebib{If not otherwise specified all values are taken from \citet{vanderhucht}; (1) \citet{huchtwr11}; (2) \citet{NorthWR}; (3) \citet{Millour2007}; (4) \citet{DeMarco2}; (5) \cite{Wallace}; (6) \cite{Williams}; (7) \cite{dougherty2}; (8) \cite{vanHuchtwr146dist}; (9) \cite{MorrisWR147dist}; (10) \cite{Marchenko}; (11) \cite{dougherty1}; (12) \cite{SetiaWR147period}; (13) \cite{Romero2003}; (14) \cite{Williams1990}; (15) \cite{SetiaWR140velocity}; (16) \cite{doughertyWR146}; (17) \cite{vacca}; (18) \cite{Willis}; (19) \cite{crowtherTemp}; (20) \cite{wr137orbit}; (21) \cite{monnier}; omitted (...) values are not known.}
\end{table*}

\end{document}